%% file: arxiv3.tex
\documentclass[prl,twocolumn,showpacs]{revtex4-1}
\usepackage{amsmath,mathrsfs,graphicx}

\newcount\minute
\newcount\hour
\newcount\hourMins
\def\now
{
  \minute=\time
  \hour=\time \divide \hour by 60
  \hourMins=\hour \multiply\hourMins by 60
  \advance\minute by -\hourMins
  \zeroPadTwo{\the\hour}:\zeroPadTwo{\the\minute}
}\def\timestamp
{
  \today\ \now
}
\def\today
{
  \zeroPadTwo{\the\day}-\zeroPadTwo{\the\month}-\the\year%
}
\def\zeroPadTwo#1%
{% Left zero pad of the argument to 2 digits.  The argument
  \ifnum #1<10 0\fi
  #1%
}
\def \dif {\mathrm{d}}
\def \Dif {\mathcal{D}}

\date{\timestamp}
\begin{document}

\title{Phase transition of interacting disordered bosons in one dimension}

\author{Zoran Ristivojevic,$^1$ Aleksandra Petkovi\'{c},$^1$ Pierre Le Doussal,$^1$ and Thierry Giamarchi$^2$}

\affiliation{$^1$Laboratoire de Physique Th\'{e}orique--CNRS, Ecole Normale Sup\'{e}rieure, 24 rue Lhomond, 75005 Paris, France}
\affiliation{$^2$DPMC-MaNEP, University of Geneva, 24 Quai Ernest-Ansermet, CH-1211 Geneva, Switzerland}

\begin{abstract}
Interacting bosons generically form a superfluid state. In the presence of disorder it can get converted into a compressible Bose glass state. Here we study such transition in one dimension at moderate interaction using bosonization and renormalization group techniques. We derive the two-loop scaling equations and discuss the phase diagram. We find that the correlation functions at the transition are characterized by universal exponents in a finite region around the fixed point.
\end{abstract}

\maketitle

Disordered quantum systems show a remarkable richness of phenomena. One of the most spectacular manifestations of the interplay between quantum effects and disorder is provided
by Anderson localization, where the interference between waves can lead to exponentially localized wave functions. For noninteracting particles, Anderson localization
is by now well understood and has been spectacularly tested in various experiments ranging from microwaves to cold atoms \cite{Aspect+09}. The situation is considerably more complex when interactions are present, and a full understanding of disordered interacting quantum systems is largely still missing. Among such systems, particularly interesting are disordered bosons. Indeed in the absence of disorder, interacting bosons are expected to undergo a Bose-Einstein condensation and to become superfluid (SF), and hence naively should be able to resist disorder. One can thus expect the competition between disorder and interactions to be particularly relevant in that case.
In addition since all noninteracting bosons collapse into a disorder created lowest energy state, interactions must be included from the start to understand this competition.

Indeed it was shown by a renormalization group (RG) analysis \cite{Giamarchi+88} that disordered interacting bosons would undergo a phase transition between a SF
and a localized phase. This transition and this phase, nicknamed Bose glass (BG), was also shown \cite{Fisher+89} by scaling arguments to exist in higher dimensions.
In one dimension the SF-BG transition is in the Berezinskii-Kosterlitz-Thouless (BKT) universality class, and universal exponents at the transition exist
for the various correlation functions, in particular the single particle one. One interesting possibility, which was argued in \cite{Giamarchi+88}
was the possibility of the existence of two localized phases, the BG corresponding to a strong interaction, strong disorder fixed point, while another would
correspond to a weak interaction, strong disorder fixed point. These predictions on the phase diagram and the flow were found to be in good agreement with numerical
studies of this problem \cite{Prokofev+98PhysRevLett.80.4355,Rapsch+99}.

Recently this problem has regained a considerable interest thanks to its remarkable realization in cold atomic gases \cite{Billy+08,Roati+08,Pasienski+10}. On the theoretical side, new studies focused on the low interaction, strong disorder case \cite{Lugan+07PhysRevLett.98.170403,Fontalesi+11PhysRevA.83.033626,vosk+12PhysRevB.85.024531,Nattermann}. In particular using a real-space renormalization group study of a related disordered Josephson junction array model a BKT-like transition was found again, but now the Luttinger parameter takes disorder dependent value at the transition \cite{Altman+10}. These results are thus in contradiction with the universal exponent found by the RG analysis \cite{Giamarchi+88} for intermediate interactions, which would then suggest the existence of two distinct phases. It is thus important to ascertain that the universality of the exponent found in \cite{Giamarchi+88} is not
an artefact of the use of the lowest order RG but survives if higher order are taken into account. This is important in view of what can happen for other disordered
systems such as the Cardy-Ostlund model \cite{Cardy+82} for which some nonuniversal terms are generated at next order. In addition pushing the RG flow to the next
order is also useful in view of the comparison with e.g.~numerical studies of the phase diagram.

In this letter we perform an analysis of the SF-BG transition using a Luttinger liquid (LL) description of the interacting bosons and find
an RG flow to second order in disorder. We show, both from the RG flow and from symmetry arguments that the exponents remain universal at the transition whenever the LL description
can be used. This implies that the exponents will stay universal on at least a part of the phase diagram. If there are indeed varying exponents at weak interactions
a transition between two localized phases must exist. We also clarify the respective influences of the forward and backward scattering on the impurities and compute
the single particle and density correlation functions.

We describe a quantum interacting Bose fluid in one dimension.
Its low energy properties can be described by the Tomonaga Luttinger liquid model \cite{Giamarchi} which corresponds to the Hamiltonian
\begin{align} \label{eq:llham}
{H}_0=\frac{\hbar}{2\pi}\int\dif x\left\{ vK [\partial_x{{\theta}}(x)]^2+\frac{v}{K}[\partial_x{{\varphi}}(x)]^2 \right\},
\end{align}
where $v$ denotes the sound velocity and $K$ is a dimensionless Luttinger parameter. The fields $\theta$ and $\varphi$ satisfy $[\theta(x),\partial_y\varphi(y)]=i\pi\delta(x-y)$. The density of particles reads
%\begin{align}\label{density}
$
\rho(x)=\rho_0+2\rho_1\partial_x\varphi(x)+2\rho_2\cos[2\varphi(x)-2\pi\rho_0x],
$
%\end{align}
where $\rho_1=-1/2\pi$, $\rho_0$ is the average density, and $\rho_2$ is a nonuniversal constant. For a local repulsion $g$ between bosons, one has $0<1/K<1$ for $0<g<\infty$. \cite{footnote1}

In addition we put the system in a disordered potential:
\begin{align}
H_{d}= \int\dif x\left\{2\rho_1\eta(x)\partial_x\varphi+\rho_2\left[\xi^*(x) \mathrm{e}^{i2\varphi}+\text{h.c.}\right]\right\},
\end{align}
where we distinguish the forward $\eta$ and backward $\xi$ scattering part of disorder with Fourier components around momenta $0$ and $\pm 2\pi\rho_0$, respectively \cite{Giamarchi}. We assume Gaussian disorder with correlations
$
\overline{\eta(x)\eta(x')}=\hbar^2D_f\delta(x-x'),\quad \overline{\xi(x)\xi^*(x')}=\hbar^2D_b\delta(x-x'),
$
while the other correlations vanish. By $\overline{.\phantom{1}.\phantom{1}.}$ we denote the disorder average. The total Hamiltonian of the system is $H=H_0+H_{d}$.

We average over disorder using the replica trick. The replicated Euclidean action is $S=S_0+S_f+S_b$. The quadratic part reads
\begin{align}\label{S0}
\frac{S_0}{\hbar}=&\frac{v}{2\pi K} \int_{x\tau\atop\alpha}\Big[(\partial_x\varphi_\alpha)^2 +\frac{1}{v^2}(\partial_\tau\varphi_{\alpha})^2+ m^2(\varphi_\alpha)^2\Big],
\end{align}
where $\varphi_\alpha(x,\tau)$ is a set of bosonic fields. Here $\alpha = 1\ldots n$ and the limit $n\to 0$ has to be taken in the end.
We introduced the shorthand notation $\int_{x\tau\atop\alpha}\ldots\equiv\sum_{\alpha}\int\dif x\dif\tau\ldots$, where Greek letters denote replica indices.
As an infrared cutoff we introduced a small mass $m$ which will be sent to zero at the end of calculations. The disorder part of the replicated action reads
\begin{align}\label{Sf}
&\frac{S_f}{\hbar}=-2\rho_1^2D_f\int_{x\tau\tau'\atop{\alpha\beta}} [\partial_x\varphi_\alpha(x,\tau)] [\partial_x\varphi_\beta(x,\tau')],\\
\label{Sb}
&\frac{S_b}{\hbar}=-\rho_2^2D_{b} \int_{x\tau\tau'\atop{\alpha\beta}} \cos[2\varphi_\alpha(x,\tau)-2\varphi_\beta(x,\tau')].
\end{align}

Introducing the Fourier transform one can diagonalize
the harmonic terms (\ref{S0}) and (\ref{Sf}) in $S$. The correlation function $G_{\alpha\beta}(x,\tau)=\langle \varphi_\alpha(x,\tau)\varphi_\beta(0,0)\rangle$ is then written \cite{footnote2} as $G_{\alpha\beta}=G\delta_{\alpha\beta}+G_f$, with
\begin{align}\label{G}
&G(x,\tau)=\frac{K}{2}K_0(m\sqrt{x^2+v^2\tau^2+a^2}),\\
\label{G0}
&G_f(x)=\frac{\pi^2 K^2}{v^2}\frac{\rho_1^2D_f}{m}\mathrm{e}^{-m|x|}(1-m|x|).
\end{align}
Here the parameter $a$ has been introduced as the ultraviolet cutoff, and $\langle\ldots\rangle$ is the average with respect to the harmonic part $(S_0+S_f)/\hbar$. $K_0$ denotes the modified Bessel function of the second kind \cite{Abramowitz}.

To treat the backward scattering part of the disorder, we use a field theoretic approach \cite{Zinn-Justin,Amit+80,Ristivojevic+12} to obtain the effective action $\Gamma$ of the model. For an action $S(\varphi)$, it is defined as $\Gamma(\varphi)=J\varphi-W(J)$, where $W(J)$ is the generator of connected correlations given by
$
{{W}(J)}=\ln\int\Dif\varphi \mathrm{e}^{-S(\varphi)/\hbar+J\varphi}.
$
Using $J(x)=\frac{\delta \Gamma}{\delta \varphi(x)}$,
we obtain \cite{Zinn-Justin}
%\begin{align}\label{Gamma}
$
{\Gamma(\varphi)}=-\ln\int\Dif\chi \mathrm{e}^{-S(\varphi+\chi)/\hbar+\int\dif x\chi(x)\frac{\delta \Gamma}{\delta \varphi(x)}}.
$
%\end{align}
This general equation can be solved perturbatively, order by order with respect to the small parameter $D_b$ \cite{Ristivojevic+12}.
The perturbative expansion of the effective action contains all the information about critical properties of our model (at the first two orders in $D_b$, in our case). In order to derive the scaling equations one should find the divergent terms in $\Gamma$ in the limit $a\to 0$. After expanding the operators one finally obtains the effective action \cite{footnote3}
\begin{widetext}
\vspace{-0.5cm}
\begin{align}\label{Gammafinal}
\Gamma=&\sum_{\alpha} \int\dif x\dif\tau\left\{\frac{v}{2\pi K}\left[(\partial_x\varphi_\alpha)^2+m^2(\varphi_\alpha)^2\right]+ \left[\frac{1}{2\pi K v}+2\mathscr{B}a_1+\mathcal{O}(\mathscr{B}^2)\right] (\partial_\tau\varphi_\alpha)^2\right\} -\left[2\rho_1^2D_f+2\mathscr{B}^2b_2\right]\notag\\
&\times\sum_{\alpha\beta}\int\dif x\dif\tau\dif\tau'[\partial_x\varphi_\alpha(x,\tau)] [\partial_x\varphi_\beta(x,\tau')]
-\left[\mathscr{B}+2b_1\mathscr{B}^2\right]\sum_{\alpha\beta}\int\dif x\dif\tau\dif\tau' \cos[2\varphi_\alpha(x,\tau)-2\varphi_\beta(x,\tau')],
\end{align}
\vspace{-0.3cm}
\end{widetext}
where we have introduced $\mathscr{B}=\rho_2^2D_b\mathrm{e}^{-4G(0,0)}$.
The coefficients $a_1,b_1$, and $b_2$ read
$a_1=\int\dif\tau\tau^2\left[\mathrm{e}^{4G(0,\tau)}-1\right]$,
$b_2=\int\dif x\dif\tau\dif\tau'x^2 [\mathrm{e}^{4G(x,\tau)}-1][\mathrm{e}^{4G(x,\tau')}-1]$, and
$b_1=\int\dif x\dif\tau\dif\tau'f_2(x,\tau,\tau+\tau')$,
where
$
f_2(x,\tau,\tau')=\mathrm{e}^{4G(x,\tau)-4G(x,\tau')+4G(0,\tau-\tau')} -\mathrm{e}^{4G(x,\tau)}-\mathrm{e}^{-4G(x,\tau')}
-\mathrm{e}^{4G(0,\tau-\tau')} +2+4[\mathrm{e}^{4G(0,\tau-\tau')}-1][G(x,\tau')-G(x,\tau)]
$.

We can now calculate the unknown parameters in (\ref{Gammafinal}). Introducing the small parameter  $\delta=K-3/2$ which measures the distance from the critical point (see below) and using (\ref{G}) we get
\begin{align}\label{a1}
a_1=\frac{-2\lambda+ \delta\lambda^2-4(\ln2-1)\delta\lambda+\mathcal{O}(\delta^2) +2c_1}{2(cmv)^3},
\end{align}
where $\lambda=\ln c^2m^2a^2$, $c=\mathrm{e}^{\gamma_E}/2$,  $\gamma_E$ is the Euler constant, and $c_1$ is a constant.
The next two terms are
\begin{align}\label{b1}
b_1 &=\frac{\pi\left[9\lambda^2+2(51-54\ln2)\lambda +\mathcal{O}(\delta)+4c_2\right]}{8v^2(cm)^3},\\
b_2 &=\frac{2\pi}{v^2(cm)^6}\frac{1}{a}\left[1+\mathcal{O}(\delta)\right]. \label{b2}
\end{align}

The effective action (\ref{Gammafinal}) has divergencies  when $ma\to0$ contained in $\lambda$. In order to remove them we introduce a set of renormalized coupling constants (denoted by the subscript $_R$) by $\mathscr{D}=Z_b \mathscr{D}_R,v=Z v_R,\delta=Z(3/2+\delta_R)-3/2,K=Z(3/2+\delta_R),m_R=m$
where
\begin{align}
Z_b=&1-\frac{1}{2}\left[\left(39-54\ln 2+9c_1\right)\mathscr{D}_R+2\delta_R\right]\lambda\notag\\
&+\frac{1}{4}\left(9\mathscr{D}_R+2\delta_R^2\right)\lambda^2 -\mathscr{D}_R(6c_1+c_2),\\
Z=&1-\left[3\mathscr{D}_R+\left(6\ln 2-4\right)\mathscr{D}_R\delta_R\right]\lambda +\frac{3}{2}\mathscr{D}_R\delta_R\lambda^2\notag\\
&+3c_1\mathscr{D}_R+2c_1\mathscr{D}_R\delta_R.
\end{align}
We introduced the dimensionless disorder strength $\mathscr{D}=\pi \rho_2^2 a^3D_b/v^2$. The effective action expressed in terms of renormalized quantities does not contain divergencies. This leads to the RG equations of the coupling constants, obtained by requiring that derivatives of the initial coupling constants with respect to $m_R$ nullify.
They read
\begin{align}\label{D}
&\frac{\dif\mathscr{D}_R}{\dif\ell}=-2\mathscr{D}_R\delta_R+A\mathscr{D}_R^2 +\mathcal{O}(\mathscr{D}_R^2\delta_R),\\
\label{delta}
&\frac{\dif\delta_R}{\dif\ell}=-9\mathscr{D}_R+B\mathscr{D}_R\delta_R +\mathcal{O}(\mathscr{D}_R^2),\\
\label{Df}
&\frac{\dif }{\dif\ell}(aD_{fR})=0,\quad \frac{\dif}{\dif\ell}\left(\frac{v_R}{K_R}\right)=0,
\end{align}
where $\ell=-\ln m_R$. The constants  $A=54\ln 2-39-9c_1$ and $B=6-18\ln2+9c_1$ depend on the details of the choice of regularization in (\ref{S0}) \cite{footnote4}. However, their sum
\begin{align}
\label{invariant}
A+B=36\ln2-33
\end{align}
is a universal number for the model. Equations (\ref{D})-(\ref{invariant}) are the main results of this letter.

\begin{figure}
\includegraphics[width=0.8\columnwidth]{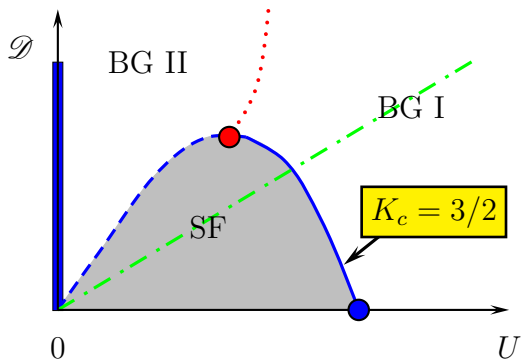}
\caption{\label{fig1} (color online)
Possible phase diagram of a disordered 1D Bose gas as a function of the boson repulsion $U$ and disorder $\mathscr{D}$. The green dash-dotteed line schematically indicates the boundary below which a bosonization description of such a problem is guaranteed (namely the disorder is smaller than the chemical potential). In this region
the superfluid-Bose glass transition, denoted by the solid blue line,
is described by \emph{universal} exponents. The question of the nature and critical behavior of the transition in the regime for which bosonization cannot be directly applied is yet open (see text). One possibility (not shown on the figure) is that the exponent remains universal along the whole SF-BG line, leading to a single BG phase. The only singular line is then the noninteracting bosons ($U=0$) which are localized by rare events of the disorder and indicated by the blue box.
The other possibility (shown in the figure), is that if at small interaction, there are nonuniversal exponents along the dashed blue line,
then this forces the existence of a critical point on the SF-BG boundary between the nonuniversal and universal regime. This would imply the existence of two distinct BG phases.}
\end{figure}

At the first order in the backward scattering (i.e., putting $A=B=0$) the above equations are identical to those first derived in \citet{Giamarchi+88} (see also \cite{Giamarchi}). One could notice the absence of renormalization to $v/K$ which an exact result in all orders due to the statistical symmetry $\varphi(x,\tau)\to\varphi(x,\tau)+w(x)$ of the disordered part of the action \cite{Schulz+88,Hwa+94}.
The second order terms are novel and lead to additional insight in the physics of the SF-BG transition for this model. At the fist order in $D_b$ no \textit{off-diagonal} (in replica space) renormalization of the quadratic action could be generated, for trivial reasons. The second order is the lowest order at which such terms might in principle appear. From (\ref{Gammafinal}) we see that no such terms are generated, except for terms describing the forward scattering process, which have no consequences for the localization properties of the system. In addition, the forward scattering terms contain no divergence and hence there is no renormalization of $D_f$. The quadratic part of the effective action thus remains, even to the order $D_b^2$ essentially diagonal in replica. The transition between the SF and the BG phase can thus be fully characterized by the two parameters $K$ and $D_f$, and not, like in other models such as the classical Cardy-Ostlund model \cite{Cardy+82}, by a full set of variables (corresponding to the off-diagonal terms) that enter into the correlation functions and can affect the exponents of correlation functions at the transition. This important result, proven here directly from the RG flow, is a consequence of the time independence of the disorder \cite{Giamarchi+96PhysRevB.53.15206}. Indeed for two independent replicas $\alpha\neq\beta$ \emph{before} averaging over disorder one obtains
\begin{align} \label{eq:aver}
\langle \varphi_\alpha(x,\tau) \varphi_\beta(x',\tau') \rangle = &\langle \varphi_\alpha(x,\tau) \rangle \langle \varphi_\beta(x',\tau') \rangle\notag\\
=&\langle \varphi_\alpha(x,0) \rangle \langle \varphi_\beta(x',0) \rangle
\end{align}
since the disorder does not depend on $\tau$. The correlation (\ref{eq:aver}) is thus time independent.
Thus terms such as $\int \dif x\dif x' \dif\tau [\partial_\tau \varphi_\alpha(x,\tau)][\partial_\tau \varphi_\beta(x',\tau)]$ cannot
appear in the effective action, at any order, since they would lead to time dependence for (\ref{eq:aver}).
This has important consequences for the physical properties at the separatrix between the phase for which the disorder is irrelevant ($D_b \to 0$) and the phase for which the disorder is relevant (the BG phase). The absence of such off-diagonal replica terms thus leads
to a \emph{universal} value of the parameter $K$, namely $K_c=3/2$, and correlation functions will thus decay with a \emph{universal} exponent at the transition. Our analysis thus confirms that the SF-BG transition around the value $K=3/2$, i.e. for intermediate interactions, has a generic \emph{universal} exponent in a finite region around that point. This puts stringent constraints on the phase diagram, as schematized in Fig.~\ref{fig1}.
Indeed this result that uses the bosonization representation of the disordered Bose gas is guaranteed to work when the disorder is weaker than the chemical potential.
This means that for a good part of the SF-BG boundary the exponent will remain constant. What happens for larger disorder or weaker interactions is still an open question.
One possibility is of course that the exponent remains universal along the whole line. In that case, there is most likely a unique BG phase. The only singular line
of the phase diagram would be in that case, the non-interacting line $U=0$ for which the bosons are localized by rare events of the random potential in a \emph{finite}
region of space. However if for weak interactions one can obtain non-universal exponents at the transition SF-BG as discussed in \cite{Altman+10}, it is then impossible
to smoothly connect the two separatrices between the SF and the BG. It would thus imply that there is on the separatrix a critical point above which the exponents would
be universal and below which they would vary with parameters. This would have for consequence to have two different localized BG phases,
as was argued as one of the possible scenarios in \cite{Giamarchi+88}. A possible order parameter distinguishing these two localized phases might be the moments of SF stiffness distributions. An interesting question
is whether a similar mechanism occurs inside the superfluid as well (see e.g.~\cite{Gurarie+08PhysRevLett.101.170407}).

The SF-BG transition can be seen directly from the superfluid correlation function, which will go from a divergent power law behavior
in the SF phase, to an exponentially decreasing one in the localized phase. Varying the strength of the disorder and of the interactions
allows to probe the universality of the exponent at the transition. Such behavior can be probed in cold atomic systems or in magnetic insulators \cite{giamarchi+08naturephysics, Zheludev+10PhysRevB.81.060410, Yamada+11PhysRevB.83.020409,Zapf+11arxiv}. Cold atoms offer the advantage of the control over disorder and interactions. However optical lattice systems suffer from the inhomogeneity due to the confining potential,
which complicates the analysis of the exponent. Atom chips realizations are up to now limited to relatively small interactions ($K\sim 40$).
Magnetic insulators are very homogeneous and allow precise control and measurement of the boson density and compressibility. Controlling the disorder is, however, more challenging. No doubt that for both types of systems these difficulties will be overcome in the future, allowing reliable
answers to the above points.

Let us now use our improved RG flow to investigate more quantitatively the behavior close to the transition.
The solution of the flow equations (\ref{D}) and (\ref{delta}) is
\begin{align}\label{solution}
\delta_R^2-9\mathscr{D}_R-A\delta_R\mathscr{D}_R+2(A+B)\delta_R^3/27=C,
\end{align}
where $C$ is an arbitrary constant, and $C<0$ ($C>0$) marks the insulating (SF) phase. For $C=0$ the system is at the critical line, where $\mathscr{D}_R=(\delta_R/3)^2+(2B-A)\delta_R^3/243+\mathcal{O}(\delta_R^4)$. Then we obtain the solution of (\ref{delta}) at large scales
\begin{align}\label{flowdeltal}
&\delta_R(\ell)\simeq \frac{1}{\ell}+\frac{(A+B)}{27}\frac{\ln\ell}{\ell^2}.
\end{align}
In general the solution of RG equations (\ref{solution}) is nonuniversal. However along the critical line $\delta_R(\ell)$ is \textit{universal} since the invariant of the model (\ref{invariant}) emerges, which in turn determines universal form of the correlation functions at the transition, as shown below. Since $\mathscr{D}_R\sim\delta_R^2$ close to the transition the neglected term $\mathcal{O}(\mathscr{B}^2)$ in (\ref{Gammafinal}) would produce $\mathcal{O}(\mathscr{D}_R^2)$ in (\ref{delta}) which is beyond the two loop order we consider. The correlation length close to the critical line from the insulating side ($C<0$) takes the form $\xi\propto \exp(\pi/\sqrt{9\mathscr{D}_R-\delta_R^2})[1+\mathcal{O}(\delta_R)]$. Therefore higher order corrections do not affect $\xi$ in an essential way and the transition is of BKT type \cite{Giamarchi+88}.

The two loop RG also allows us to compute the single particle correlation function $F(x)=\langle\Psi(x)\Psi^\dagger(0)\rangle_H$ at the transition with improved accuracy. It can be obtained from the Euclidean action $S$ as $F(x)=\rho_0
\langle\mathrm{e}^{i\left[\theta(x,0)-\theta(0,0)\right]}\rangle_S$. Performing the perturbation theory with respect to (\ref{Sb}) and taking into account the flow of parameters one obtains for $|x|\gg a$ \begin{align}\label{correlationf1}
\!\!\!{F}(x)=\rho_0\left(\frac{a}{|x|}\right)^{1/3}\left(\ln\frac{|x|}{a}\right)^{2/9} \left[1-\frac{2(A+B)}{243}\frac{\ln\ln\frac{|x|}{a}}{\ln\frac{|x|}{a}}\right]\!.
\end{align}
Note that forward scattering term does not appear in $F(x)$.
In addition to the lowest order result ${F(x)}\propto |x|^{-1/3}$ \cite{Giamarchi+88}, we find a logarithmic correction $\ln^{2/9}(|x|/a)$. Finally the role of the universal quantity (\ref{invariant}) is found in subleading corrections to $F(x)$ in (\ref{correlationf1}).

The $2\pi\rho_0$ part of the density-density correlation function $\mathcal{F}(x,\tau)=\langle\mathrm{e}^{i2\varphi(x,\tau)} \mathrm{e}^{-i2\varphi(0,0)}\rangle_S$ can be obtained in a similar way. On the separatrix it reads
\begin{align}\label{correlationf}
\mathcal{F}(x,\tau)\propto & \exp\left[-8\pi^2 \frac{K^2}{v^2} \rho_1^2D_f|x|\right] \left(\frac{a}{R}\right)^{3}\left({\ln\frac{R}{a}}\right)^{-2}\notag\\ &\times\left[1+\frac{2(A+B)}{27}\frac{\ln\ln\frac{R}{a}}{\ln\frac{R}{a}}\right],
\end{align}
where $R=\sqrt{x^2+v_R^2\tau^2}\gg a$. The first term of (\ref{correlationf}) is due to the forward scattering term, and we see that is comes with $D_f K^2/v^2$ that is not renormalized.

In conclusion, we have computed to two loop order the RG equations for interacting disordered bosons.  We have shown, both from the RG and from general arguments that in the regime described by bosonization (intermediate interactions) the separatrix between the BG and SF is characterized by a \emph{universal exponent}. We computed the single particle and density-density correlations at the transition. Our calculation uses bosonization and therefore also applies to disordered fermions. Detailed aspects of that and the details of calculation will be published elsewhere.

We acknowledge discussions with E. Altman, G. Refael, and B. Svistunov. The work at ENS is supported by the ANR Grant No.~09-BLAN-0097-01/2. This work was supported in part by the Swiss NSF under MaNEP and Division II.

%\bibliography{bibliography-GS}

%merlin.mbs apsrev4-1.bst 2010-07-25 4.21a (PWD, AO, DPC) hacked
%Control: key (0)
%Control: author (8) initials jnrlst
%Control: editor formatted (1) identically to author
%Control: production of article title (-1) disabled
%Control: page (0) single
%Control: year (1) truncated
%Control: production of eprint (0) enabled
%

%\newpage
%\includepdf[pages={{},1,{},2}]{boseglass-PRL-resubmit-supplement.pdf}
%
\clearpage

\input{arxiv3-supplement.tex}

\end{document}

%% file: arxiv3-supplement.tex
\onecolumngrid
\begin{center}{\large
\textbf{Phase transition of interacting disordered bosons in one dimension\\
--Supplemental material--}
}
\end{center}
\vskip 5mm
\begin{center}
Zoran Ristivojevic,$^1$ Aleksandra Petkovi\'{c},$^1$ Pierre Le Doussal,$^1$ and Thierry Giamarchi$^2$
\end{center}

\begin{center}\textit{
$^1$Laboratoire de Physique Th\'{e}orique--CNRS, Ecole Normale Sup\'{e}rieure, 24 rue Lhomond, 75005 Paris, France and\\
$^2$DPMC-MaNEP, University of Geneva, 24 Quai Ernest-Ansermet, CH-1211 Geneva, Switzerland
}
\end{center}

\vskip 5mm

\onecolumngrid
\footnotetext

The defining equation for the effective action from the main text reads
\begin{align}
{\Gamma(\varphi)}=-\ln\int\Dif\chi \mathrm{e}^{-S(\varphi+\chi)/\hbar+\int\dif x\chi(x)\frac{\delta \Gamma}{\delta \varphi(x)}}.
\end{align}
This general equation can be solved perturbatively, order by order with respect to the small parameter $D_b$ \cite{Ristivojevic+12supp}.
Up to an additive constant, we obtain $\Gamma=(S_0+S_f)/\hbar+\Gamma_1+\Gamma_2+\mathcal{O}(D_b^3)$, with $\Gamma_1=\langle S_b(\varphi+\chi)\rangle_\chi/\hbar$ and
\begin{align}
\label{Gamma2def}
\Gamma_2=-\frac{\langle S_b^2(\varphi+\chi)\rangle_\chi}{2\hbar^2}+\frac{1}{2}\Gamma_1^2 +\frac{1}{2}\sum_{\alpha\beta}\int\dif x\dif\tau\dif\tau' G_{\alpha\beta}(x-x',\tau-\tau')\frac{\delta\Gamma_1}{\delta\varphi_\alpha(x,\tau)} \frac{\delta\Gamma_1}{\delta\varphi_\beta(x',\tau')},
\end{align}
where $\langle\cdots\rangle_\chi$ denotes an average with respect to the field $\chi$. From the technical point of view, our approach is very similar to the standard Wilsonian approach when one performs thinning of high-momentum degrees of freedom. In the present case the difference is that the fields $\chi$ that are integrated out live in the whole momentum space and in a cumulant expansion calculates the effective Hamiltonian. Averages that appear during such procedure of integrating out high-momentum degrees of freedom are with respect to the quadratic Hamiltonian and one uses Wick's theorem, see for example Ref.~\cite{Kogut} for details.

At the lowest order we easily obtain
\begin{align}\label{Gamma1}
\Gamma_1=&-\mathscr{B} \sum_{\alpha\beta}\int\dif x\dif \tau\dif\tau'\left\{\left[\mathrm{e}^{4G(0,\tau-\tau')}-1\right] \delta_{\alpha\beta}+1\right\}\cos[2\varphi_\alpha(x,\tau)-2\varphi_\beta(x,\tau')],
\end{align}
where $\mathscr{B}=\rho_2^2D_b\mathrm{e}^{-4G(0,0)}$.
One should notice that only a part of the full correlation function that is diagonal in replica indices, (6), enters the result. This is due to the fact that the off-diagonal part (7) of the propagator does not depend on imaginary time.

The second order term is more complicated and could be written as a sum $\Gamma_2=\sum_{j=1}^3\Gamma_2^{(j)}$, where $j$ denotes the number of replica sums in the expressions:
\begin{align}
\label{Gamma21}
\Gamma_2^{(1)}=&-\frac{1}{2}\mathscr{B}^2\int_{x\tau\tau' x_1\tau_1\tau_1'\atop\alpha} f_1(x-x_1,\tau,\tau',\tau_1,\tau_1') \cos[2\varphi_\alpha(x,\tau)-2\varphi_\alpha(x,\tau')+ 2\varphi_\alpha(x_1,\tau_1)-2\varphi_\alpha(x_1,\tau_1')],\\
%\end{align}
%\begin{align}
\label{Gamma22}
\Gamma_2^{(2)}=&-\mathscr{B}^2\sum_{s=\pm1}\int_{x\tau\tau' x_1\tau_1\tau_1'\atop{\alpha\beta}} \bigg\{f_2(x-x_1,\tau-\tau_1',\tau-\tau_1) \cos[2\varphi_\alpha(x,\tau)+2\varphi_\alpha(x_1,\tau_1) -2\varphi_\alpha(x_1,\tau_1') -2\varphi_\beta(x,\tau')]\notag\\
&\qquad\qquad+\frac{1}{2}f_3(x-x_1,\tau-\tau_1,\tau'-\tau_1',s) \cos[2\varphi_\alpha(x,\tau)-2s\varphi_\alpha(x_1,\tau_1)-2\varphi_\beta(x,\tau') +2s\varphi_\beta(x_1,\tau_1')]\bigg\},\\
%\end{align}
%\begin{align}
\label{Gamma23}
\Gamma_2^{(3)}=&-
\mathscr{B}^2\sum_{s=\pm1}\int_{x\tau\tau' x_1\tau_1\tau_1'\atop{\alpha\beta\gamma}} f_4(x-x_1,\tau-\tau_1,s) \cos[2\varphi_\alpha(x,\tau') -2\varphi_\beta(x,\tau)+2s\varphi_\beta(x_1,\tau_1) -2s\varphi_\gamma(x_1,\tau_1')].
\end{align}
The $f$ functions in the previous equations are
$
f_2(x,\tau,\tau')=\mathrm{e}^{4G(x,\tau)-4G(x,\tau')+4G(0,\tau-\tau')} -\mathrm{e}^{4G(x,\tau)}-\mathrm{e}^{-4G(x,\tau')}
-\mathrm{e}^{4G(0,\tau-\tau')} +2+4[\mathrm{e}^{4G(0,\tau-\tau')}-1][G(x,\tau')-G(x,\tau)],
$
$
f_3(x,\tau,\tau',s)=[\mathrm{e}^{4sG(x,\tau)}-1][\mathrm{e}^{4sG(x,\tau')}-1]$, $f_4(x,\tau,s)=\mathrm{e}^{4sG(x,\tau)}-4sG(x,\tau)-1$.
The expression (\ref{Gamma21}) leads to corrections of $K$ of the order $\mathcal{O}(D_b^2)$, that is a higher order in the expansion than the one considered in Eqs.~(15) and (19) of the main text.

The obtained perturbative expansion of the effective action contains all the information about critical properties of our model at the first two orders in $D_b$. In order to derive the scaling equations one should find the divergent terms in $\Gamma$ in the limit $a\to 0$. After expanding the operators one obtains the effective action
\begin{align}\label{Gammafinal-sup}
\Gamma=&\sum_{\alpha} \int\dif x\dif\tau\left\{\frac{v}{2\pi K}\left[(\partial_x\varphi_\alpha)^2+m^2(\varphi_\alpha)^2\right]+ \left[\frac{1}{2\pi K v}+2\mathscr{B}a_1+\mathcal{O}(\mathscr{B}^2)\right] (\partial_\tau\varphi_\alpha)^2\right\} -\left[2\rho_1^2D_f+2\mathscr{B}^2b_2\right]\notag\\
&\times\sum_{\alpha\beta}\int\dif x\dif\tau\dif\tau'[\partial_x\varphi_\alpha(x,\tau)] [\partial_x\varphi_\beta(x,\tau')]
-\left[\mathscr{B}+2b_1\mathscr{B}^2\right]\sum_{\alpha\beta}\int\dif x\dif\tau\dif\tau' \cos[2\varphi_\alpha(x,\tau)-2\varphi_\beta(x,\tau')].
\end{align}
The coefficients $a_1,b_1$, and $b_2$ arise from the expansion of operators in (\ref{Gamma1}) and (\ref{Gamma22}) and read
$a_1=\int\dif\tau\tau^2\left[\mathrm{e}^{4G(0,\tau)}-1\right]$, $b_1=\int\dif x\dif\tau\dif\tau'f_2(x,\tau,\tau+\tau')$, and $b_2=\int\dif x\dif\tau\dif\tau'x^2f_3(x,\tau,\tau',1)$. The terms from (\ref{Gamma22}) produce corrections to operators  (8), contained in $b_1$ and $b_2$. One can notice that due to the parity $f_3(x,\tau,\tau',1)=f_3(x,-\tau,-\tau',1)$ the analogous term to (4) but with time derivatives is \emph{absent} in (8). This differs from what happens in the the related classical 2D Cardy-Ostlund model \cite{Cardy+82supp,Ristivojevic+12supp}
where such terms are generated. The three-replica part (\ref{Gamma23}) contains two contributions that turn out to be unimportant. Namely, for $s=1$ after a gradient expansion a free sum over $\beta$ index delivers a factor $n$, that vanishes in the replica limit. For $s=-1$, one generates an operator of the form $\propto\cos(2\varphi_\alpha+2\varphi_\gamma-4\varphi_\beta)$ which has a nondivergent prefactor and in addition is irrelevant close to the critical point.

%% file: arxiv3.bbl
\begin{thebibliography}{30}
\bibitem{footnote1}{Note that the LL
  description (\ref{eq:llham}) covers the case of a fermionic fluid as well
  with proper identification of the parameters. There $v$ denotes the plasmon
  velocity, $K<1\ (K>1)$ corresponds to repulsive (attractive) interaction.}
\bibitem{footnote2}{For the matrix
  $M_{\alpha\beta}=M_1\delta_{\alpha\beta}+M_2$ we have used
  $(M_{\alpha\beta})^{-1}=\delta_{\alpha\beta}/M_1-M_2/M_1^2$ valid for $n\to
  0$.}
\bibitem{footnote4}{The constant $c_1$
  in (\ref{a1}) depends on the correlation function (\ref{G}) beyond distances
  $m^{-1}$, hence is nonuniversal.}
\end{thebibliography}
